\documentclass[12pt]{article}
\usepackage[centertags]{amsmath}
\usepackage{amsfonts}
\usepackage{amssymb}
\usepackage{amsthm} 
\usepackage{newlfont}
\usepackage{epsfig}
\usepackage{amscd}
\usepackage{graphicx}
\usepackage{epsfig}
\usepackage{amscd}

\newcommand{\beq}{\begin{equation}}
\newcommand{\eeq}{\end{equation}}
\newcommand{\ba}{\begin{array}}
\newcommand{\ea}{\end{array}}
\newcommand{\bea}{\begin{eqnarray}}
\newcommand{\eea}{\end{eqnarray}}
\begin{document}

\begin{center}
{\large \sc \bf {
Robust stationary distributed discord in  Jordan-Wigner fermion system 
 under perturbations of initial state 
}}

\vskip 15pt

{\large 
 E.B.Fel'dman and A.I.~Zenchuk 
}

\vskip 8pt

{\it Institute of Problems of Chemical Physics, RAS,
Chernogolovka, Moscow reg., 142432, Russia},\\
 e-mail:  efeldman@icp.ac.ru, zenchuk@itp.ac.ru 

\end{center}

%\today

\begin{abstract}
We investigate the Jordan-Wigner  fermion clusters 
with the stationary distributed pairwise quantum discord. Such clusters appear after the  
Jordan-Wigner transformation of  
a spin chain governed by 
the nearest-neighbor XY-Hamiltonian with  the particular initial state having one polarized node.
We show that the quantum discord stationarity  in such systems is not  
destroyed by the "parasitic" polarization of at least  two types. 
First type appears because the initial state with a single polarized node 
is hardly realizable experimentally, so that 
the low polarization of neighboring nodes  must be taken into account. 
Second, the additional noise-polarization of all nodes is unavoidable.  
Although  the stationarity  may not be destroyed  by 
 perturbations of the above two types, the parasitic polarizations 
 deform the distribution of the pairwise discord and may destroy the 
 clusters of  correlated fermions with equal pairwise discords. 
Such deformations are studied in this paper.
 \end{abstract}

%\maketitle

%%%%%%%%%%%%%%%%
\section{Introduction}
\label{Sec:Introduction}
The quantum correlations are responsible for the effective operation of  quantum information
devices having the essential advantages in comparison with their classical counterparts 
\cite{Werner,Z0,HV,OZ,Z,AFOV,HHHH,BDFMRSSW,HHHOSS,NC,DFC,DV,DSC,LBAW}.
According to the current stand point, the total correlations in 
multi-particle system are described by  mutual information 
and quantum correlations for both pure and mixed states are characterized  by the
quantum discord \cite{Z0,HV,OZ,Z,NC,DFC,DV,DSC,LBAW}. 

Studing the quantum correlations, it is important to choose the proper 
quantum system possessing the desirable properties and realizable in practice. 
In this regard we point on the chains of nuclear spins, which are  suitable 
 for  realization of  quantum registers and quantum  devices  transferring and 
manipulating the quantum information. It is challenging that  the multiple quantum (MQ) 
NMR  methods \cite{BMGP,DMF} allow us to construct experimentally the XY-interacting spin chains. 
Moreover,  using the NMR method,  it is possible to create  conditions providing the concentration of
polarization  at a single node of the chain (up to the unavoidable  experimental errors) \cite{ZME}.
The dynamic of quantum correlations in this  model was first studied in \cite{FBE}.
Moreover, it was shown recently \cite{FZ_2012} that  such chains
are  convenient  for study of the dependence of the discord on the representation basis of the 
density matrix describing the state of a quantum system.
Thus, the  quantum discord calculated for 
the interacting nuclear spins  differs from that between the
fermions arising after the Jordan-Wigner transformation \cite{JW}  
of the  density matrix operator \cite{FZ_2012,FZ_2014}. 
It seemed out that the quantum discord between fermions may exhibit very interesting 
properties  \cite{FZ_2012} which have not been observed 
in the discord  between nuclear spins. The most important property 
is the   stationarity of the pairwise  discord in a fermion cluster with 
the above initial state of a spin-1/2 chain.
Besides, if we polarize the proper initial node, then the quantum discord 
will be the same for any fermion pair in the selected fermion cluster. Apparently, this fact is important 
for implementation of the fermion registers  in the quantum devices because all
fermion-nodes are equivalent from the 
quantum correlation point of view. 

The existence of such clusters motivates the study of their 
stability with respect to both the
experimental errors in creation of the single-node polarization and the noise effects.
We  note that the stability of the spin dynamics  in the presence of different types of 
noise is an attractive problem, because noise is  unavoidable in any quantum process. 
In particular, the fidelity of the perfect (in the absence of noises) state transfer under  
noise-perturbations of the coupling constants in the Hamiltonian was considered in  
refs.\cite{CRMF,ZASO,ZASO2,ZASO3,SAOZ} for two chains: a completely
engineered chain and a chain with  remote end-nodes. In both cases, the  important result is 
that the noise reduces the fidelity without changing 
the state transfer time.

In this paper we  study  the stability of the discord  distribution relative to 
perturbations of the initial state in a homogeneous spin chain 
(i.e. the coupling constants in the Hamiltonian are assumed to be stable).  
We show that the stationarity of 
the quantum discord in the system with 
the single initially polarized node  may not be destroyed by the additional 
low polarizations of the neighboring nodes which unavoidably appear in the experiment.
This perturbation just leads to the deformation of the pairwise quantum discord and may eventually  destroy 
the clusters of fermions with the equal pairwise discord. 
The threshold value of the low-polarization is found. 
 We also consider the deformation of the stationary 
discord distribution caused by the 
noise-polarization appearing in all nodes of the spin chain. It is remarkable that the discord stationarity 
is not disturbed in both cases.

The paper is organized as follows. The Jordan-Wigner transformation of the 
$XY$-Hamiltonian with nearest neighbor interactions   is briefly 
discussed in Sec.\ref{Section:JW}. The stability of
the pairwise discord  stationarity in the 
Jordan-Wigner fermion system 
of a spin-1/2 chain with single initially polarized node under perturbations of the 
initial state   is demonstrated in Secs.\ref{Section:chain} and  \ref{Section:noise} with the numerical 
simulations of the spin dynamics of a 17-node chain. 
First, in Sec.\ref{Section:chain}, the parasitic polarization of two neighboring nodes 
(with respect to the selected inner polarized node) is considered. Then,
in Sec.\ref{Section:noise}, the noise polarization of all nodes 
is taken into account using the perturbation method. Deformations of the fermion clusters with the
equal pairwise discord   under the above perturbations are also considered in 
Secs.\ref{Section:chain} and  \ref{Section:noise}. The basic 
results are discussed in Sec.\ref{Section:conclusions}. A formula for  calculating  
the discord in the $X$-type 
density matrix \cite{ARA} is represented in the Appendix, Sec.\ref{Section:appendix}.

%%%%%%%%%%%%%%
\section{Jordan-Wigner transformation of  $XY$-Hamiltonian with the
nearest-neighbor interaction.}
\label{Section:entire}\label{Section:JW}
Hereafter  we study  quantum correlations in the one-dimensional open 
spin-1/2 chain of $N$ nodes governed by the 
XY Hamiltonian with the nearest neighbor interactions,
\begin{eqnarray}\label{XY}
H=\omega_0 \sum_{i=1}^N I_{iz} + 
D \sum_{i=1}^{N-1} (I_{ix} I_{(i+1)x} + I_{iy} I_{(i+1)y}),
\end{eqnarray}
where $\omega_0$ is the Larmor frequency in the external magnetic field, $D$ is 
the spin-spin coupling constant
between the nearest neighbors, and $I_{i\alpha}$ ($i=1,\dots,N$, $\alpha=x,y,z$) 
is the $i$th spin projection on 
the $\alpha$-axis. 
%We note that the proposed consideration holds for any Hamiltonian commuting with 
%the $z$-projection of the total spin momentum $I_z$.

Following refs.\cite{FBE,FZ_2012,FZ_2014},
we diagonalize 
 Hamiltonian (\ref{XY})  using  
 the Jordan-Wigner transformation method \cite{JW}
\begin{eqnarray}\label{Hdiag}
H=\sum_{k} \varepsilon_k \beta_k^+\beta_k -\frac{1}{2} N \omega_0,\;\;
\varepsilon_k = D \cos(k) +\omega_0,
\end{eqnarray}
where the fermion operators $\beta_j$ are determined in terms of   other fermion operators $c_j$  
by means of the Fourier transformation
\begin{eqnarray}\label{betc}
\beta_k = \sum_{j=1}^N g_k(j) c_j,
\end{eqnarray}
and the fermion operators  $c_j$ are determined as \cite{JW}
\begin{eqnarray}
c_j=(-2)^{j-1} I_{1z}I_{2z}\dots I_{(j-1)z} I^-_j.
\end{eqnarray}
Here,
\begin{eqnarray}
g_k(j)=\left(
\frac{2}{N+1}
\right)^{1/2} \sin(k j),\;\;\;\displaystyle k=\frac{\pi n}{N+1}, \;\;\;n=1,2,\dots,N.
\end{eqnarray}
%with $\displaystyle k=\frac{\pi n}{N+1}$, $n=1,2,\dots,N$.
We can readily  express the  projection operators $I_{jz}$ 
in terms of the fermion operators $c_j$ as
\begin{eqnarray}\label{Izc}
I_{jz} = c^+_j c_j -\frac{1}{2},\;\;\forall \; j.
\end{eqnarray}
Hereafter,  diagonal representation (\ref{Hdiag})  of the $XY$ Hamiltonian 
 will be used  to describe the dynamics of the density matrix associated with 
 the spin-1/2 chain.

%%%%%%%%%%%%%%%%%%%

\section{Initial state with three polarized nodes}
\label{Section:chain}
The  dynamics of the Jordan-Wigner fermions associated with  the spin-1/2 chain with
 single initially polarized node $j_0$ has been studied in refs.\cite{FBE,FZ_2012,FZ_2014}. 
 The stationarity of the pairwise  discord in such systems is demonstrated therein. Also the fermion 
 clusters with the equal pairwise discord are revealed in those references.
 
Now we consider the initial state 
 with an inner  initially polarized node $j_0$ (i.e., $1<j_0<N$) and assume the 
 parasitic low polarization of 
 two neighboring nodes, i.e. the initial density matrix reads
\begin{eqnarray}\label{instgen}
&&
\rho_0=\frac{e^{\sum_{k=j_0-1}^{j_0+1}b_k I_{k,z}}}{Z}=
\frac{1}{2^N}\prod_{k=j_0-1}^{j_0+1}
\left(1+2 I_{kz}\tanh \frac{b_k}{2}\right),\\\nonumber
&&
Z={\mbox{Tr}} (\prod_{k=j_0-1}^{j_0+1}e^{b_k I_{kz}}) = 
2^N \prod_{k=j_0-1}^{j_0+1}\cosh\frac{b_k}{2},
\end{eqnarray}
where $b_j=\frac{\hbar \omega_{j0}}{kT}$,   $\hbar$ is the Plank constant, 
$k$ is the Boltzmann constant, and $T$ is the temperature of the system. 

The motivation for considering this initial state is mentioned in the introduction. Namely, 
an experimental scheme may not provide the ideal single-node polarization. 
So, two neighboring nodes $j_0\pm 1$  get some polarization as well provided that  $j_0$ is the inner node,
i.e. $1<j_0<N$. 
This polarization  might
be referred to as the parasitic one. 
As was shown in \cite{FZ_2012,FZ_2014}, a fermion cluster with equal pairwise discords 
(which is of our interest in this paper) may be obtained if the polarized node $j_0$ 
is an inner one. 
This case is related to  the density matrix (\ref{instgen}) and  is considered below.

The evolution of the initial density matrix (\ref{instgen}) in virtue of the fermion representation of the 
Hamiltonian (\ref{Hdiag}) reads
\begin{eqnarray}\label{rhoev}
\rho(t)=e^{-i t \sum_k \varepsilon_k \beta^+_k\beta_k} \rho_0
e^{i t \sum_k \varepsilon_k \beta^+_k\beta_k}.
\end{eqnarray}
Using the  identity
\begin{eqnarray}\label{identity}
e^{-i \varphi \beta^+_k \beta_k} \beta^+_k e^{i \varphi \beta^+_k \beta_k} = e^{-i \varphi} \beta^+_k,\;\;
\forall \;\varphi,
\end{eqnarray}
we transfer 
density matrix (\ref{rhoev})
to the following form \cite{FBE}
\begin{eqnarray}\label{rhot30}\label{rhot0e}
&&
\rho(t)=\frac{1}{2^N} \prod_{j=j_0-1}^{j_0+1} 
\left(1-\tanh\frac{b_j}{2}+ 2\tanh\frac{b_j}{2} \sum_{k,k'}
e^{-i t(\varepsilon_k-\varepsilon_{k'})} g_k(j) g_{k'}(j) 
\beta^+_k\beta_{k'}\right)=\\\nonumber
&&
A^{j_0}_0+
 \sum_{k,k'}A^{j_0}_{kk'} e^{-i t(\varepsilon_k-\varepsilon_{k'})} \beta^+_k\beta_{k'}+
\sum_{k,k',q,q'} 
A^{j_0}_{kqk'q'}e^{-i t(\varepsilon_k+\varepsilon_q-\varepsilon_{k'}-\varepsilon_{q'})} 
 \beta^+_k\beta_{k'}\beta^+_q\beta_{q'}+
 \\\nonumber
 &&
 \sum_{k,k',q,q',l,l'} 
A^{j_0}_{kqlk'q'l'}e^{-i t(\varepsilon_k+\varepsilon_q+\varepsilon_l-
\varepsilon_{k'}-\varepsilon_{q'}-\varepsilon_{l'})} 
 \beta^+_k\beta_{k'}\beta^+_q\beta_{q'}\beta^+_l\beta_{l'}
\end{eqnarray} 
where
\begin{eqnarray}
&&
A^{j_0}_0=\frac{1}{2^N}
\prod_{j=j_0-1}^{j_0+1}(1-\tanh\frac{b_j}{2}),\\\nonumber
&&
A^{j_0}_{kk'}=\frac{1}{2^{N-1}} \Big(\\\nonumber
&&
(1-\tanh\frac{b_{j_0}}{2}) (1-\tanh\frac{b_{j_0+1}}{2}) \tanh\frac{b_{j_0-1}}{2}
g_k(j_0-1) g_{k'}(j_0-1) 
+\\\nonumber
&&
(1-\tanh\frac{b_{j_0-1}}{2}) (1-\tanh\frac{b_{j_0+1}}{2}) \tanh\frac{b_{j_0}}{2}
g_k(j_0) g_{k'}(j_0) 
+\\\nonumber
&&
(1-\tanh\frac{b_{j_0-1}}{2}) (1-\tanh\frac{b_{j_0}}{2}) \tanh\frac{b_{j_0+1}}{2}
g_k(j_0+1) g_{k'}(j_0+1) \Big),
\\\nonumber
 &&
 A^{j_0}_{kqk'q'}=\frac{1}{2^{N-2}} \Big( \\\nonumber
&&
(1-\tanh\frac{b_{j_0-1}}{2}) 
\tanh\frac{b_{j_0}}{2} \tanh\frac{b_{j_0+1}}{2} 
g_k(j_0) g_{k'}(j_0) g_q(j_0+1) g_{q'}(j_0+1)
+\\\nonumber
&&
(1-\tanh\frac{b_{j_0}}{2}) 
\tanh\frac{b_{j_0-1}}{2} \tanh\frac{b_{j_0+1}}{2} 
g_k(j_0-1) g_{k'}(j_0-1) g_q(j_0+1) g_{q'}(j_0+1)
+\\\nonumber
&&
(1-\tanh\frac{b_{j_0+1}}{2}) 
\tanh\frac{b_{j_0-1}}{2} \tanh\frac{b_{j_0}}{2} 
g_k(j_0-1) g_{k'}(j_0-1) g_q(j_0) g_{q'}(j_0)\Big),\\\nonumber
&&
A^{j_0}_{kqlk'q'l'} =\\\nonumber
&&
 \frac{1}{2^{N-3}} \tanh\frac{b_{j_0-1}}{2}\tanh\frac{b_{j_0}}{2} \tanh\frac{b_{j_0+1}}{2} 
g_k(j_0-1) g_{k'}(j_0-1)
g_q(j_0) g_{q'}(j_0) g_l(j_0+1) g_{l'}(j_0+1)
\end{eqnarray}
Finally, using the fermion anti-commutation
relations \cite{LL} and  relations among  the coefficients $A^{j_0}_{kqk'q'}$ and $A^{j_0}_{kqlk'q'l'}$
\begin{eqnarray}\label{Asym}
\sum_{q=1}^N A^{j0}_{kqqk'} = \sum_{q=1}^N A^{j0}_{klqk'ql'}= \sum_{q=1}^N A^{j0}_{kqlqk'l'}=
\sum_{q=1}^N A^{j0}_{klqqk'l'}=0,
\end{eqnarray}
which follow from the orthogonality relationship 
\begin{eqnarray}
\sum_{k=1}^N g_k(j) g_k(l)=\delta_{kl},
\end{eqnarray}
we easily transform eq.(\ref{rhot0e}) to the following canonic form:
\begin{eqnarray}\label{rhot3}
&&
\rho(t)=\\\nonumber
&&
\frac{1}{2^N}
\prod_{j=j_0-1}^{j_0+1}(1-\tanh\frac{b_j}{2})+
 \sum_{k,k'} A^{j_0}_{kk'}e^{-i t(\varepsilon_k-\varepsilon_{k'})} \beta^+_k\beta_{k'}-
\\\nonumber
&&
\sum_{{k,k',q,q'}\atop{k\neq q, k'\neq q'}} 
 A^{j_0}_{kqk'q'}e^{-i t(\varepsilon_k+\varepsilon_q-\varepsilon_{k'}-\varepsilon_{q'})}
 \beta^+_k\beta^+_{q}\beta_{k'}\beta_{q'}-
 \\\nonumber
&&
 \sum_{{k,k',q,q',l,l'}\atop{k\neq q\neq l, k'\neq q'\neq l'}} 
A^{j_0}_{kqlk'q'l'}e^{-i t(\varepsilon_k+\varepsilon_q+\varepsilon_l-
\varepsilon_{k'}-\varepsilon_{q'}-\varepsilon_{l'})}
 \beta^+_k\beta^+_q\beta^+_l\beta_{k'}\beta_{q'}\beta_{l'}.
\end{eqnarray} 

%%%%%%%%%%%%%%%%%%%%%%%%%
\subsection{Reduced density matrix}
\label{Section:reduced}
Studing  quantum correlations we consider only the pairwise discord. First, 
in calculation of
the discord between the $n$th and $m$th fermions, we have to 
 reduce  density matrix (\ref{rhot3}) with respect to all 
nodes except for the $n$th and $m$th ones resulting to the marginal density matrix of the
following form:
\begin{eqnarray}\label{rednm2}
\rho_{nm}^{j_0}(t)=
B^{j_0}_{nm} + \sum_{k,k'=n,m} B^{j_0}_{nmkk'} 
e^{-i t(\varepsilon_k-\varepsilon_{k'})}\beta^+_k\beta_{k'} 
+  C^{j_0}_{nm}\beta^+_n\beta^+_m\beta_{m}\beta_n 
,
\end{eqnarray}
where all coefficients do not depend on the  time $t$:
\begin{eqnarray}
&&
B^{j_0}_{nm} = \frac{1}{4} \prod_{j=j_0-1}^{j_0+1} 
\left(1-\tanh\frac{\beta_j}{2}\right)+2^{N-3} \sum_{k\neq n,m}  A^{j_0}_{kk}+
2^{N-4}
\sum_{{k,q}\atop{k\neq q;k,q\neq n,m}} 
(- A^{j_0}_{kqkq}+ A^{j_0}_{kqqk})-\\\nonumber
 &&
2^{N-5}\sum_{{k,q,l\neq n,m}\atop{k\neq q\neq l}} 
(-A^{j_0}_{kqlkql}+A^{j_0}_{kqlklq}+A^{j_0}_{kqlqkl}-A^{j_0}_{kqllkq} - 
A^{j_0}_{kqlqlk}+ 
A^{j_0}_{kqllqk}
 )
\\\nonumber
&&
 B^{j_0}_{nmkk'} =2^{N-2} A^{j_0}_{kk'} +
 2^{N-3}
 \sum_{q,q\neq k,k'}(- A^{j_0}_{kqk'q} + A^{j_0}_{kqqk'} +
 A^{j_0}_{qkk'q}-
 A^{j_0}_{qkqk'})
 -\\\nonumber
 &&
 2^{N-4}\sum_{{q,l}\atop{ q\neq l\neq k \neq k'}} 
(-A^{j_0}_{kqlk'ql}+A^{j_0}_{kqlk'lq}+A^{j_0}_{kqlqk'l}-A^{j_0}_{kqllk'q} - A^{j_0}_{kqlqlk'}+
A^{j_0}_{kqllqk'}+
A^{j_0}_{qklk'ql}-
\\\nonumber
&&
A^{j_0}_{qklk'lq}-A^{j_0}_{qklqk'l}+A^{j_0}_{qkllk'q}    
+A^{j_0}_{qklqlk'}-A^{j_0}_{qkllqk'} -
\\\nonumber
&&
A^{j_0}_{qlkk'ql} +A^{j_0}_{qlkk'lq} +A^{j_0}_{qlkqk'l} -A^{j_0}_{qlklk'q}-
A^{j_0}_{qlkqlk'} +A^{j_0}_{qlklqk'} 
)
 \\\nonumber
 &&
C^{j_0}_{nm}  =  
2^{N-2} ( A^{j_0}_{nmmn} -  A^{j_0}_{nmnm} -A^{j_0}_{mnmn} + A^{j_0}_{mnnm})-
 \\\nonumber
 &&
 2^{N-3}\sum_{k\neq n,m} (
A^{j_0}_{knmkmn} -
A^{j_0}_{knmknm}-
A^{j_0}_{kmnkmn} + 
A^{j_0}_{kmnknm} - \\\nonumber
 &&
A^{j_0}_{knmmkn}+
A^{j_0}_{knmnkm}+
A^{j_0}_{kmnmkn}-
A^{j_0}_{kmnnkm}+
\\\nonumber
 &&
A^{j_0}_{knmmnk}-
A^{j_0}_{knmnmk}-
A^{j_0}_{kmnmnk}+
A^{j_0}_{kmnnmk}-
A^{j_0}_{nkmkmn} + 
A^{j_0}_{nkmknm}+
A^{j_0}_{mknkmn} - 
A^{j_0}_{mknknm} + 
\\\nonumber
 &&
A^{j_0}_{nkmmkn}-
A^{j_0}_{nkmnkm}-
A^{j_0}_{mknmkn}+
A^{j_0}_{mknnkm}-
A^{j_0}_{nkmmnk}+
A^{j_0}_{nkmnmk}+
A^{j_0}_{mknmnk}-
A^{j_0}_{mknnmk}+
\\\nonumber
 &&
A^{j_0}_{nmkkmn} -
A^{j_0}_{nmkknm}-
A^{j_0}_{mnkkmn} + 
A^{j_0}_{mnkknm} - 
A^{j_0}_{nmkmkn}+
A^{j_0}_{nmknkm}+
A^{j_0}_{mnkmkn}-
A^{j_0}_{mnknkm}+
\\\nonumber
 &&
A^{j_0}_{nmkmnk}-
A^{j_0}_{nmknmk}-
A^{j_0}_{mnkmnk}+
A^{j_0}_{mnknmk})
\end{eqnarray}
Next, using the basis 
\begin{eqnarray}\label{basis}
|00\rangle,\;|01\rangle,|10\rangle,\;|11\rangle,
\end{eqnarray}
we represent the  marginal density matrix operator (\ref{rednm2})  in the following  matrix form:
\begin{eqnarray}\label{rednm3}
\rho^{j_0}_{nm}(t)=
\left(
\begin{array}{cccc}
B^{j_0}_{nm} &0&0&0\cr
0&B^{j_0}_{nm} +B^{j_0}_{nmnn}&B^{j_0}_{nmnm}e^{-i t(\varepsilon_n-\varepsilon_{m})}&0\cr
0&B^{j_0}_{nmmn}e^{i t(\varepsilon_n-\varepsilon_{m})}&B^{j_0}_{nm} +B^{j_0}_{nmmm}&0\cr
0&0&0&B^{j_0}_{nm} +B^{j_0}_{nmmm}+B^{j_0}_{nmnn} + C^{j_0}_{nm} 
\end{array}
\right).
\end{eqnarray}

\subsubsection{Stationarity of the pairwise discord}
\label{Section:stationarity}
Formula (\ref{rednm3}) shows that the 
diagonal elements of marginal   matrix (\ref{rednm3}) do not depend on 
the time $t$. The $t$-dependence appears only
in  non-diagonal elements.
However, the pairwise discord for the $X$-matrix depends on the absolute value $|B^{j_0}_{nmnm}|$ 
of the non-diagonal element 
(see the Appendix, Sec.\ref{Section:appendix})
and, consequently, 
does not depend on $t$. This means that the perturbations 
considered in this section 
do not destroy the stationarity of the discord. However, the distribution of the discord becomes deformed
which, eventually, may destroy the fermion clusters with equal pairwise discords. 
Deformations caused by the parasitic polarizations 
of $(j_0\pm 1)$th nodes are studied 
numerically in the following subsection.

%%%%%%%%%%%%%%%%%%%%
\subsection{Numerical simulations}
\label{Section:num1}
We represent  numerical simulations for a particular case of  $N=17$
node spin chain and assume that the polarization is initially concentrated at such a node $j_0$
that the fermion clusters with  equal pairwise discord may be selected from the whole system of 17 fermions
\cite{FZ_2014}. 
The interest in this case appears because such clusters might be promising in the QIP-devices as 
candidates for large quantum registers.

In accordance with \cite{FZ_2012,FZ_2014}, such a cluster $Cl$ appears in an odd-node spin chain
in two cases. First, 
if $j_0$ is the middle
node  ($j_0=9$ in our case), then the cluster $Cl$ is formed by the odd-fermions. Second,  
if $N=5+6 i$ ($i=1,2,\dots$) and $j_0=2(i+1)$ (in our case $N=17$, $i=2$, 
so that $j_0=6$), then the cluster $Cl$ is formed by the all fermions except  each third one. 
In both cases, the pairwise discord $Q_{nm}$ between the $n$th and $m$th node is
\begin{eqnarray}\label{Qnm}
Q_{nm}=\left\{
\begin{array}{ll}
Q_0=const,& n\in Cl, \;m\in Cl\cr
0,& n\notin Cl \;\;{\mbox{and/or}}\;\; m\notin Cl.
\end{array}
\right.
\end{eqnarray}
In other words,  the discord $Q_{nm}$ is zero if at least one of the subscripts $n$ 
or $m$ is not in the set $Cl$.

Thus, we consider two clusters corresponding to two   cases of the initially polarized node $j_0$:
\begin{enumerate}
%\item
%$j_0=1$, the bell-shaped cluster of all fermions.
\item
$j_0=6$, the cluster of fermions with equal pairwise discords is formed by 
 all fermions except  each third one:
 \begin{eqnarray}
\label{Cl6}
Cl = \{1,2,4,5,7,8,10,11,13,14,16,17\}.
\end{eqnarray}
\item
$j_0=9$, the  odd fermions form the cluster with equal pairwise discords:
\begin{eqnarray}
\label{Cl9}
Cl = \{1,3,5,\dots,17\}.
\end{eqnarray}
\end{enumerate}

We characterize the polarization  by the inverse temperatures 
\begin{eqnarray}
&&
b_{j_0+1} = b_{j_0-1} = b, \;\; 0 \le b \le b_{j_0},\\\nonumber
&&
b_j=0,\;\;j\neq j_0,j_0\pm 1.
\end{eqnarray}
In practice, $b$ must be such that $\tanh\frac{b_{j_0\pm 1}}{2}$ is 
several times less then
$\tanh\frac{b_{j_0}}{2}$. 
Below we  take $b_{j_0}=10$ (the low temperature limit). 

\begin{figure*}
   \epsfig{file=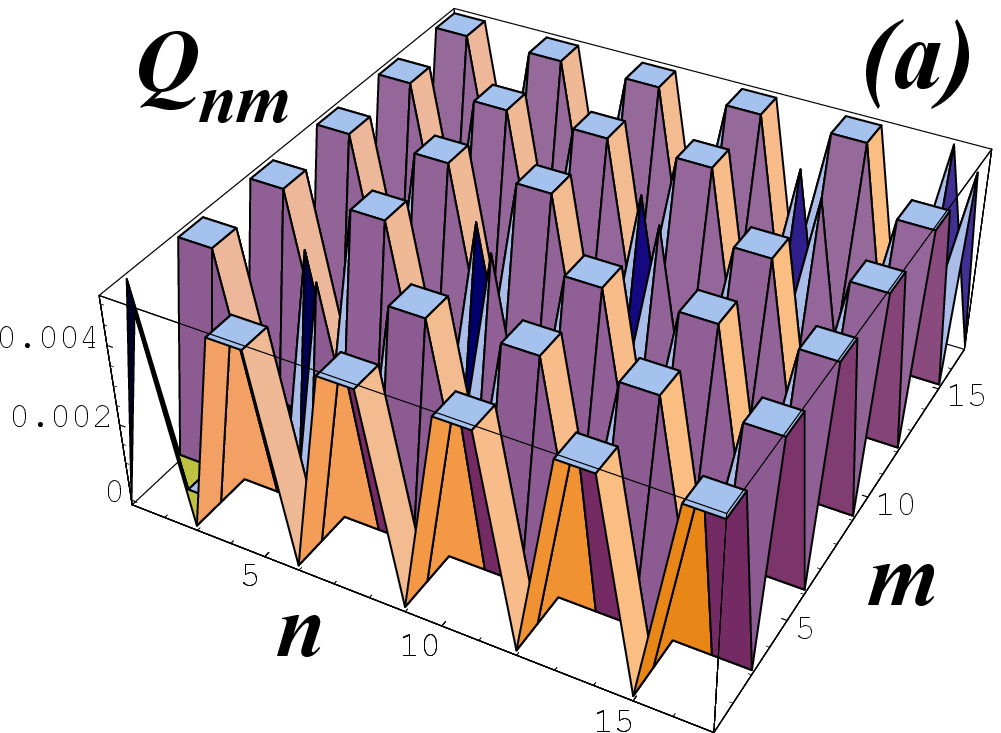,
  scale=0.8
   ,angle=0
}
\epsfig{file=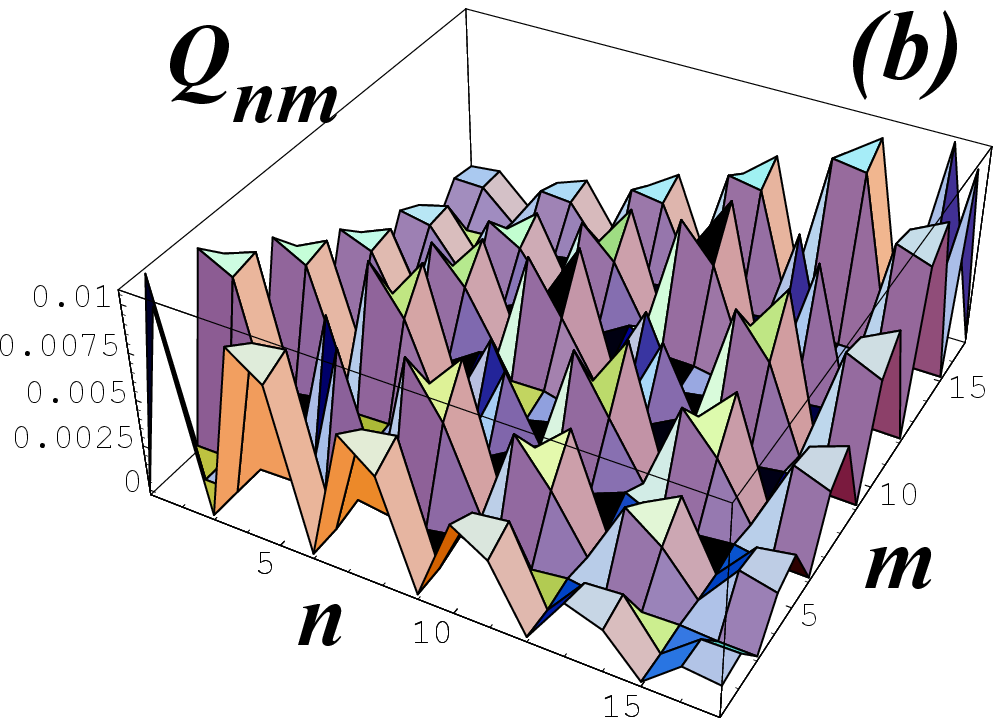, 
    scale=0.8
   ,angle=0
}
  \caption{Distribution of the discord $Q_{nm}$ in the  system of $N=17$ fermions with 
  the initially polarized node 
  $j_0=6$. Here and in Fig. \ref{Fig:3Dnoise} we put $Q_{nn}\equiv 0$ following  ref.\cite{FZ_2014}.
  ($a$) Discord $Q_{nm}$ in the fermion system without  parasitic polarization, $b=0$;
  ($b$)  Discord $Q_{nm}$ in the fermion system with the critical parasitic polarization 
  of the 5th and 7th nodes,
  $b=b_{cl}=0.480$.
} 
  \label{Fig:3Dgen} 
\end{figure*}

The presence of the parasitic polarization leads to a 
deformation of the ideal (when $b_{j_0\pm 1}=0$) discord distribution shown in Fig.\ref{Fig:3Dgen}a.
As a result, some spread of discord appears in the cluster $Cl$. Besides, the zero-valued 
discords at $b=0$ become non-zero for $b>0$.

\begin{figure*}
   \epsfig{file=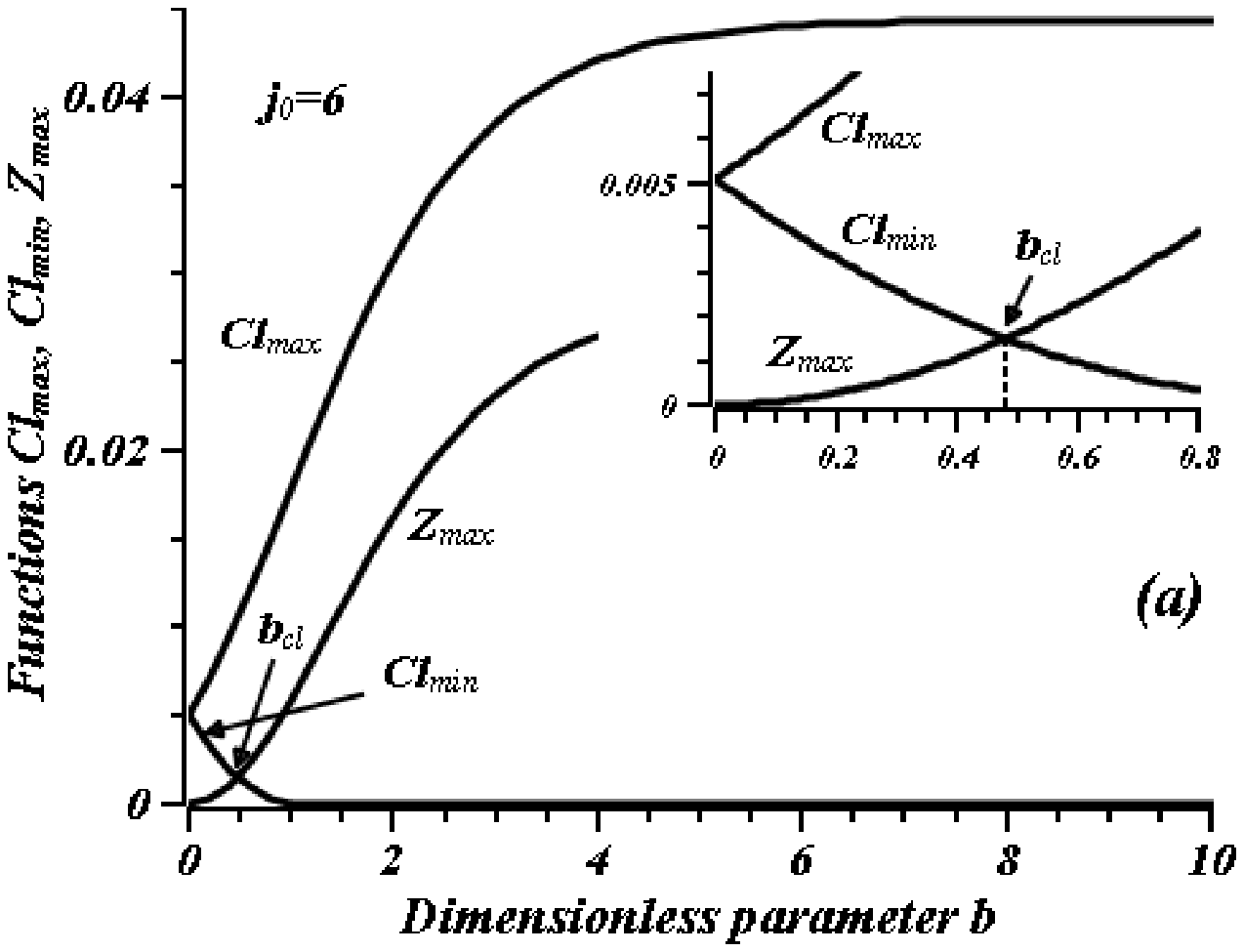,
  scale=0.6
   ,angle=0
}
\epsfig{file=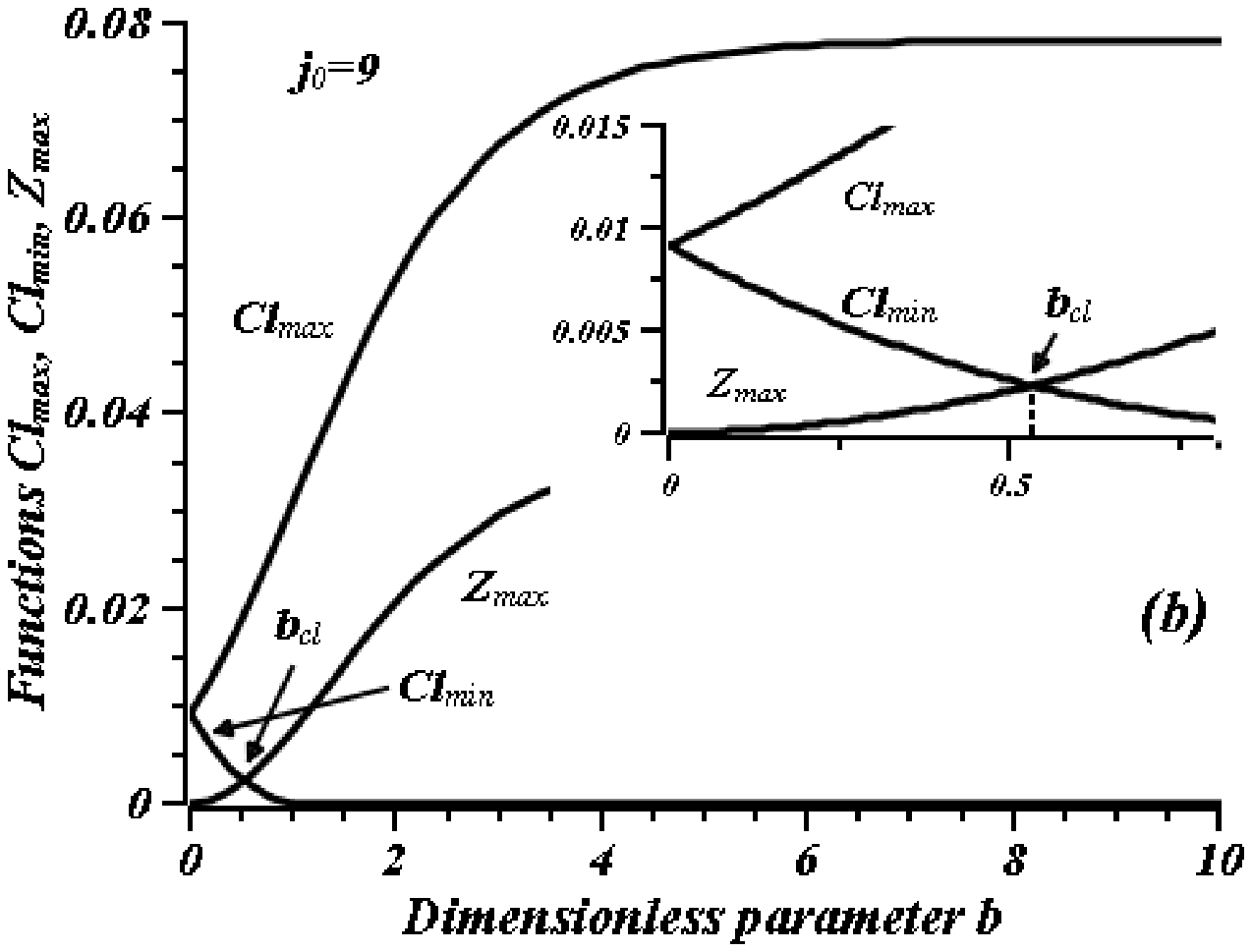, 
    scale=0.6
   ,angle=0
}
  \caption{Deformations of  fermion cluster in the system of $N=17$ fermions. 
  The graphs of the functions $Cl_{max}(b)$, 
$Cl_{min}(b)$ and $Z_{max}(b)$ are shown. The intersection point $b_{cl}$
 of the graphs
$Cl_{min}(b)$ and $Z_{max}(b)$ can be considered as the critical value of $b$
such that the cluster $Cl$ of the correlated fermions does not exist if $b>b_{cl}$. The insets show the graphs 
of $Cl_{max}(b)$, 
$Cl_{min}(b)$ and $Z_{max}(b)$ for small values of   $b$.
($a$) The initially polarized node $j_0=6$,  the critical value of the parameter $b$ is 
$b_{cl}=0.480$.
($b$) The initially polarized node $j_0=9$, the critical value  $b_{cl}=0.533$.} 
  \label{Fig:gen} 
\end{figure*}

To characterize both these effects of  parasitic polarization, 
we introduce the following functions
\begin{eqnarray}
\label{ClCl}
&&
Cl_{max}(b) =\max_{(n,m)\in Cl} {Q_{nm}}(b),\;\;Cl_{min}(b) =\min_{(n,m)\in Cl} {Q_{nm}}(b),
\\\label{ZZ}
&&
Z_{max}(b) =\max_{(n,m)\notin Cl} {Q_{nm}}(b),\;\;Z_{min}(b) =\min_{(n,m)\notin Cl} {Q_{nm}}(b).
\end{eqnarray}
Here, the notation ${(n,m)\notin Cl}$ means  $n\notin Cl\;\;{\mbox{and/or}} \;\; m\notin Cl$ .
Functions $Cl_{max}$ and  $Cl_{min}$ characterize the spread of 
the pairwise discord  in  clusters (\ref{Cl6}) and (\ref{Cl9}), while
$Z_{max}$ and  $Z_{min}$ characterize the spread of the "parasitic" discord that  was zero 
in the unperturbed case, i.e.
\begin{eqnarray}\label{cldef}
&&
Cl_{min}(b) \le Q_{nm} \le Cl_{max}(b),\;\; n\in Cl,\;m\in Cl,\\\nonumber
&&
Z_{min}(b) \le Q_{nm} \le Z_{max}(b),\;\; n\notin Cl\;\;{\mbox{and/or}}\;\;
m\notin Cl.\\\nonumber
\end{eqnarray}
The graphs of the functions $Cl_{max}(b)$, 
$Cl_{min}(b)$ and $Z_{max}(b)$  for $j_0=6$ and $j_0=9$ are shown in 
Fig.\ref{Fig:gen}a and Fig.\ref{Fig:gen}b respectively. 
The  function $Z_{min}(b)$ (although it is non-zero for $b>0$)
is not shown because it is not important in this section. But it will be used in 
Sec.\ref{Section:num2} to characterize the noise effects. 

A natural question is about the critical value of
the parameter $b$ (characterizing the  value of the "parasitic" polarization) that still does not  
completely destroy the cluster $Cl$.
We considered the value of $b$ corresponding to the  
 intersection point of 
$Cl_{min}(b)$ and $Z_{max}(b)$  as the critical  value  $b_{cl}$,
so that the cluster $Cl$  does not exist if $b>b_{cl}$. 
The critical  value is $b_{cl}=0.480$ for $j_0=6$
and $b_{cl}=0.533$ for $j_0=9$, as shown in Fig.\ref{Fig:gen}a,b.

An example of the discord distribution for $j_0=6$  and 
 $b=b_{cl}=0.48$ (at the 
 threshold value) is shown in Fig.\ref{Fig:3Dgen}b.
 We see that this distribution significantly  differs from the unperturbed case $b=0$ shown in 
 Fig.\ref{Fig:3Dgen}a. However, we emphasize ones again, the parasitic polarization does not lead to the 
 evolution of the discord in the considered fermion system, i.e.,
 the stationarity of the pairwise discord is not destroyed.

%\newpage
%%%%%%%%%%%%%%%
\section{Noise effect on the pairwise discord distribution}
\label{Section:noise}
In this section,
we assume that there is no parasitic polarization considered in Sec.\ref{Section:chain}
(i.e. $b_{j_0\pm1}=0$), but there is noise-polarization of all 
nodes. 
%and 
%the  parameters $b_j$ in eq.(\ref{rhot0}) are noised as follows
The initial density matrix (\ref{instgen}) must be replaced with the following one:
\begin{eqnarray}\label{instgen_noise}
\rho_0=
\frac{1}{2^N}\left(1+I_{j_0z} (2 \tanh \frac{b_{j_0}}{2} + \epsilon\tilde b_{j_0})  \right)
\prod_{{k=1}\atop{k\neq j_0}}^{N}
\left(1+ I_{kz} \epsilon\tilde b_k \right),\;\;Z=
2^N \cosh\frac{b_{j_0}}{2},
\end{eqnarray}
where $\epsilon$ is the double noise-polarization amplitude in a spin-1/2 chain, 
$\epsilon\ll 1$, and $\tilde b_j$ are the random numbers inside of the interval 
$-\frac{1}{2} \le \tilde b_j  \le \frac{1}{2}$. 
Using eqs.(\ref{betc}), (\ref{Izc}) %and (\ref{identity})
we transform  the initial density matrix 
to the following form \cite{FBE}
\begin{eqnarray}\label{rhot0e_noise}
&&
\rho_0=\frac{1}{2^N}
\left(a^{j_0}_0+   
\sum_{k,k'} a^{j_0}_{kk'}
 \beta^+_k\beta_{k'}\right)
 %\times\\\nonumber
%&&
\prod_{{j=1}\atop{j\neq j_0}}^N 
\left( a^{j}_0+   \sum_{k,k'}a^{j}_{kk'}
 \beta^+_k\beta_{k'}\right),
\end{eqnarray}
where
\begin{eqnarray}
&&
a^{j}_0 =\left\{
\begin{array}{ll}
1-\tanh\frac{b_{j_0}}{2} -\frac{\epsilon \tilde b_{j_0}}{2} ,& j=j_0\cr
1-\frac{\epsilon \tilde b_{j}}{2} ,& j\neq j_0\end{array}\right.,
\\\nonumber
&&
a^{j}_{kk'}= \left\{
\begin{array}{ll}(2\tanh\frac{b_{j_0}}{2} +\epsilon \tilde b_{j_0})
 g_k(j_0) g_{k'}(j_0) ,&j=j_0\cr
\epsilon \tilde b_{j}
 g_k(j) g_{k'}(j),& j\neq j_0\end{array}\right..
\end{eqnarray}
The quantities $a^{j}_{kk'}$, $j\neq j_0$, are proportional to $\epsilon$ and thus they are considered as 
small parameters hereafter.

Now we 
assume that the noise
effect can be studied by the perturbation method for small $\epsilon$. 
Thus, we 
expand  the initial density matrix  in the $a^{j}_{kk'}$, $j\neq j_0$. 
We consider two density matrices
cutting the
series and  keeping the terms up to the order, respectively,
$a^{j}_{kk'}$ and $(a^{j}_{kk'})^2$.
Taking into account the normalization 
condition  (trace of density matrix equals unit) 
%and introducing notations
%\begin{eqnarray}
%\tilde a=\prod_{j=1}^N a^{j}_0,\;\;\tilde a^{m}_l=\prod_{{j=1}\atop{j\neq l}}^N a^{j}_0,\;\;
%\tilde a_{ln}=\prod_{{j=1}\atop{j\neq l\neq m}}^N a^{j}_0,\;\;
%\tilde a_{lnm}=\prod_{{j=1}\atop{j\neq l\neq n\neq m}}^N a^{j}_0,
%\end{eqnarray}
we write these matrices as follows:
\iffalse
\begin{eqnarray}\label{rhotn}
&&
\rho_{0i}=\frac{\tilde \rho_{0i}}{ Z_i} ,\;\;Z_i=Tr\;\tilde \rho_{0i}, \;\;i=1,2
,\\\nonumber
&&
\tilde\rho_{01}=\frac{1}{2^N}\left(a^{j_0}_0+ \sum_{k,k'} a^{j_0}_{kk'} \beta^+_k\beta_{k'}\right)
\left(
\tilde a_{j_0}  +
\sum_{{n=1}\atop{n\neq j_0}}^N \tilde a_{j_0n} 
\sum_{k,k'} a^{n}_{kk'} \beta^+_k\beta_{k'} \right),
\\\nonumber
&&
\tilde\rho_{02}=\frac{1}{2^N}\left(a^{j_0}_0+ \sum_{k,k'} a^{j_0}_{kk'} \beta^+_k\beta_{k'}\right)\times\\\nonumber
&&\left(
\tilde a_{j_0}  +
\sum_{{n=1}\atop{n\neq j_0}}^N \tilde a_{j_0n} 
\sum_{k,k'} a^{n}_{kk'} \beta^+_k\beta_{k'} +
\sum_{{n,n'=1}\atop{n'\neq n\neq j_0}}^N \tilde a_{j_0nn'} 
\sum_{k,k',q,q'} a^{n}_{kk'}a^{n'}_{qq'} \beta^+_k\beta_{k'}\beta^+_q\beta_{q'}\right).
\end{eqnarray}
\fi
%\iffalse
\begin{eqnarray}\label{rhotn}
&&
\rho_{0i}=\frac{\tilde \rho_{0i}}{ Z_i} ,\;\;Z_i=Tr\;\tilde \rho_{0i}, \;\;i=1,2
,\\\nonumber
&&
\tilde\rho_{01}=\frac{1}{2^N}\left(a^{j_0}_0+ \sum_{k,k'} a^{j_0}_{kk'} \beta^+_k\beta_{k'}\right)
\left(
\prod_{{l=1}\atop{l\neq j_0}}^Na^{l}_0  +
\sum_{{n=1}\atop{n\neq j_0}}^N \left(\prod_{{m=1}\atop{m\neq j_0 \neq n}}^N a^{m}_0 \right)
\sum_{k,k'} a^{n}_{kk'} \beta^+_k\beta_{k'} \right),
\\\nonumber
&&
\tilde\rho_{02}=\frac{1}{2^N}\left(a^{j_0}_0+ \sum_{k,k'} a^{j_0}_{kk'} \beta^+_k\beta_{k'}\right)\times\\\nonumber
&&\left(
\prod_{{l=1}\atop{l\neq j_0}}^Na^{l}_0  +
\sum_{{n=1}\atop{n\neq j_0}}^N \left(\prod_{{m=1}\atop{m\neq j_0 \neq n}}^N a^{m}_0 \right)
\sum_{k,k'} a^{n}_{kk'} \beta^+_k\beta_{k'} +
\sum_{{n,n'=1}\atop{n'\neq n\neq j_0}}^N \left(\prod_{{m=1}\atop{m\neq j_0 \neq n\neq n'}}^N a^{m}_0 \right)
\sum_{k,k',q,q'} a^{n}_{kk'}a^{n'}_{qq'} \beta^+_k\beta_{k'}\beta^+_q\beta_{q'}\right).
\end{eqnarray}
%\fi
The evolution of these matrices $e^{-it H} \rho_{0i} e^{it H}$,  taking 
into account the fermion representation of the Hamiltonian (given by formula (\ref{Hdiag})) and 
  relation (\ref{identity}), reads:
 \iffalse
 \begin{eqnarray}\label{rhotnev}
&&
\rho_{i}(t)=\frac{\tilde \rho_{i}(t)}{ Z_i} ,
\\\nonumber
&&
\tilde\rho_{1}(t)=\frac{1}{2^N}\left(a^{j_0}_0+ \sum_{k,k'} a^{j_0}_{kk'}
e^{-i t (\varepsilon_k-\varepsilon_{k'})}
 \beta^+_k\beta_{k'}\right)
 %\times\\\nonumber
%&&
\left(
\tilde a_{j_0}  +
\sum_{{n=1}\atop{n\neq j_0}}^N \tilde a_{j_0n} 
\sum_{k,k'} a^{n}_{kk'} e^{-i t (\varepsilon_k-\varepsilon_{k'})}\beta^+_k\beta_{k'} \right),
\\\nonumber
&&
\tilde\rho_{2}(t)=\frac{1}{2^N}\left(a^{j_0}_0+ \sum_{k,k'} a^{j_0}_{kk'}
e^{-i t (\varepsilon_k-\varepsilon_{k'})} \beta^+_k\beta_{k'}\right)
%\times\\\nonumber
%&&
\left(
\tilde a_{j_0}  +
\sum_{{n=1}\atop{n\neq j_0}}^N \tilde a_{j_0n} 
\sum_{k,k'} a^{n}_{kk'} e^{-i t (\varepsilon_k-\varepsilon_{k'})}\beta^+_k\beta_{k'} +
\right.\\\nonumber
&&\left.
\sum_{{n,n'=1}\atop{n'\neq n\neq j_0}}^N \tilde a_{j_0nn'} 
\sum_{k,k',q,q'} a^{n}_{kk'}a^{n'}_{qq'}e^{-i t (\varepsilon_k+\varepsilon_q-
\varepsilon_{k'}-\varepsilon_{q'})} \beta^+_k\beta_{k'}\beta^+_q\beta_{q'}\right).
\end{eqnarray}
  \fi
%  
%\iffalse
  \begin{eqnarray}\label{rhotnev}
&&
\rho_{i}(t)=\frac{\tilde \rho_{i}(t)}{ Z_i} 
,\\\nonumber
&&
\tilde\rho_{1}(t)=\frac{1}{2^N}\left(a^{j_0}_0+ \sum_{k,k'} a^{j_0}_{kk'}
e^{-i t (\varepsilon_k-\varepsilon_{k'})}
 \beta^+_k\beta_{k'}\right)\times\\\nonumber
&&\left(
\prod_{{l=1}\atop{l\neq j_0}}^Na^{l}_0  +
\sum_{{n=1}\atop{n\neq j_0}}^N \left(\prod_{{m=1}\atop{m\neq j_0 \neq n}}^N a^{m}_0 \right)
\sum_{k,k'} a^{n}_{kk'} e^{-i t (\varepsilon_k-\varepsilon_{k'})}\beta^+_k\beta_{k'} \right),
\\\nonumber
&&
\tilde\rho_{2}(t)=\frac{1}{2^N}\left(a^{j_0}_0+ \sum_{k,k'} a^{j_0}_{kk'}
e^{-i t (\varepsilon_k-\varepsilon_{k'})} \beta^+_k\beta_{k'}\right)\times\\\nonumber
&&\left(
\prod_{{l=1}\atop{l\neq j_0}}^Na^{l}_0  +
\sum_{{n=1}\atop{n\neq j_0}}^N \left(\prod_{{m=1}\atop{m\neq j_0 \neq n}}^N a^{m}_0 \right)
\sum_{k,k'} a^{n}_{kk'} e^{-i t (\varepsilon_k-\varepsilon_{k'})}\beta^+_k\beta_{k'} +
\right.\\\nonumber
&&\left.
\sum_{{n,n'=1}\atop{n'\neq n\neq j_0}}^N \left(\prod_{{m=1}\atop{m\neq j_0 \neq n\neq n'}}^N a^{m}_0 \right)
\sum_{k,k',q,q'} a^{n}_{kk'}a^{n'}_{qq'}e^{-i t (\varepsilon_k+\varepsilon_q-
\varepsilon_{k'}-\varepsilon_{q'})} \beta^+_k\beta_{k'}\beta^+_q\beta_{q'}\right).
\end{eqnarray}
%\fi
Formulas (\ref{rhotnev}) may be transformed to  form (\ref{rhot30}) with
\iffalse
\begin{eqnarray}
&&
A^{j_0}_0=\frac{1}{2^N} \tilde a,\\\nonumber
&&
A^{j_0}_{kk'}= \frac{1}{2^N}\left( a^{j_0}_0
\sum_{{n=1}\atop{n\neq j_0}}^N \tilde a_{j_0n} 
 a^{n}_{kk'} + a^{j_0}_{kk'} \tilde a_{j_0}\right),\\\nonumber
 &&
A^{j_0}_{kqk'q'}=\frac{a^{j_0}_{kk'}}{2^N}
 \sum_{{n=1}\atop{n\neq j_0}}^N \tilde a_{j_0n} a^{n}_{qq'} 
,\\\nonumber
 &&
 A^{j_0}_{kqlk'q'l'}=0
\end{eqnarray}
\fi
%\iffalse
\begin{eqnarray}
&&
A^{j_0}_0=\frac{1}{2^N} \prod_{l=1}^N a_0^l,\\\nonumber
&&
A^{j_0}_{kk'}= \frac{1}{2^N}\left( a^{j_0}_0
\sum_{{n=1}\atop{n\neq j_0}}^N \left( \prod_{{m=1}\atop{m\neq j_0 \neq n}}^N a^{m}_0 \right)
 a^{n}_{kk'} + a^{j_0}_{kk'} \prod_{{l=1}\atop{l\neq j_0}}^Na^{l}_0\right),\\\nonumber
 &&
A^{j_0}_{kqk'q'}=\frac{a^{j_0}_{kk'}}{2^N}
 \sum_{{n=1}\atop{n\neq j_0}}^N \left(\prod_{{m=1}\atop{m\neq j_0 \neq n}}^N a^{m}_0 \right)a^{n}_{qq'} 
,\\\nonumber
 &&
 A^{j_0}_{kqlk'q'l'}=0
\end{eqnarray}
%\fi
for the density matrix $\rho_1$, or
\iffalse
\begin{eqnarray}
&&
A^{j_0}_0=\frac{1}{2^N} \tilde a,\\\nonumber
&&
A^{j_0}_{kk'}= \frac{1}{2^N}\left( a^{j_0}_0
\sum_{{n=1}\atop{n\neq j_0}}^N \tilde a_{j_0n} 
 a^{n}_{kk'} + a^{j_0}_{kk'} \tilde a_{j_0}\right),\\\nonumber
 &&
A^{j_0}_{kqk'q'}=\frac{1}{2^N}\left(
a^{j_0}_{kk'} \sum_{{n=1}\atop{n\neq j_0}}^N \tilde a_{j_0n} a^{n}_{qq'} +
a^{j_0}_0\sum_{{n,n'=1}\atop{n'\neq n\neq j_0}}^N 
\tilde a_{j_0nn'} 
 a^{n}_{kk'}a^{n'}_{qq'}\right),\\\nonumber
 &&
 A^{j_0}_{kqlk'q'l'}=\frac{a^{j_0}_{kk'} }{2^N}
 \sum_{{n,n'=1}\atop{n'\neq n\neq j_0}}^N
 \tilde a_{j_0nn'} 
 a^{n}_{qq'}a^{n'}_{ll'}
\end{eqnarray}
\fi
%\iffalse
\begin{eqnarray}
&&
A^{j_0}_0=\frac{1}{2^N} \prod_{l=1}^N a_0^l,\\\nonumber
&&
A^{j_0}_{kk'}= \frac{1}{2^N}\left( a^{j_0}_0
\sum_{{n=1}\atop{n\neq j_0}}^N \left(\prod_{{m=1}\atop{m\neq j_0 \neq n}}^N a^{m}_0 \right)
 a^{n}_{kk'} + a^{j_0}_{kk'} \left(\prod_{{l=1}\atop{l\neq j_0}}^Na^{l}_0\right)\right),\\\nonumber
 &&
A^{j_0}_{kqk'q'}=\frac{1}{2^N}\left(
a^{j_0}_{kk'} \sum_{{n=1}\atop{n\neq j_0}}^N \left(\prod_{{m=1}\atop{m\neq j_0 \neq n}}^N a^{m}_0
\right) a^{n}_{qq'} +
a^{j_0}_0\sum_{{n,n'=1}\atop{n'\neq n\neq j_0}}^N
\left(\prod_{{m=1}\atop{m\neq j_0 \neq n\neq n'}}^N a^{m}_0\right) 
 a^{n}_{kk'}a^{n'}_{qq'}\right),\\\nonumber
 &&
 A^{j_0}_{kqlk'q'l'}=\frac{a^{j_0}_{kk'} }{2^N}
 \sum_{{n,n'=1}\atop{n'\neq n\neq j_0}}^N
 \left(\prod_{{m=1}\atop{m\neq j_0 \neq n\neq n'}}^N a^{m}_0 \right)
 a^{n}_{qq'}a^{n'}_{ll'}
\end{eqnarray}
%\fi
for the density matrix $\rho_2$. In this section, formulas (\ref{rhot3}-\ref{rednm3}) hold   as well.
The stationarity of the pairwise quantum discord between fermions follows from the structure of the marginal 
matrix $\rho^{j_0}_{nm}$ (\ref{rednm3}) and can be shown in a way  similar to 
that  proposed in Sec.\ref{Section:chain}.

%%%%%%%%%%%%%%%%%%%%
\subsection{Numerical simulations}
\label{Section:num2}
Similar to the numerical 
simulations of Sec.\ref{Section:num1},
we perform the numerical simulations for the particular case of  $N=17$
node spin chain with the initially polarized spins  $j_0=6$ and $j_0=9$.
For each fixed value of the small parameter $\epsilon$ in the interval $0\le \epsilon \le 0.4$
($\epsilon=0,\;0.1,\;0.2,\;0.3,\;0.4$) we average the discord  over $10^2$ realizations of the random 
set of  the parameters $\tilde b_j$, $j=1,\dots,17$,
characterizing the noise-polarization 
  of the $j$th node of the spin-1/2 chain. For the  averaged discord, 
  we use the same notation  $Q_{nm}$ in this section.

\begin{figure*}
   \epsfig{file=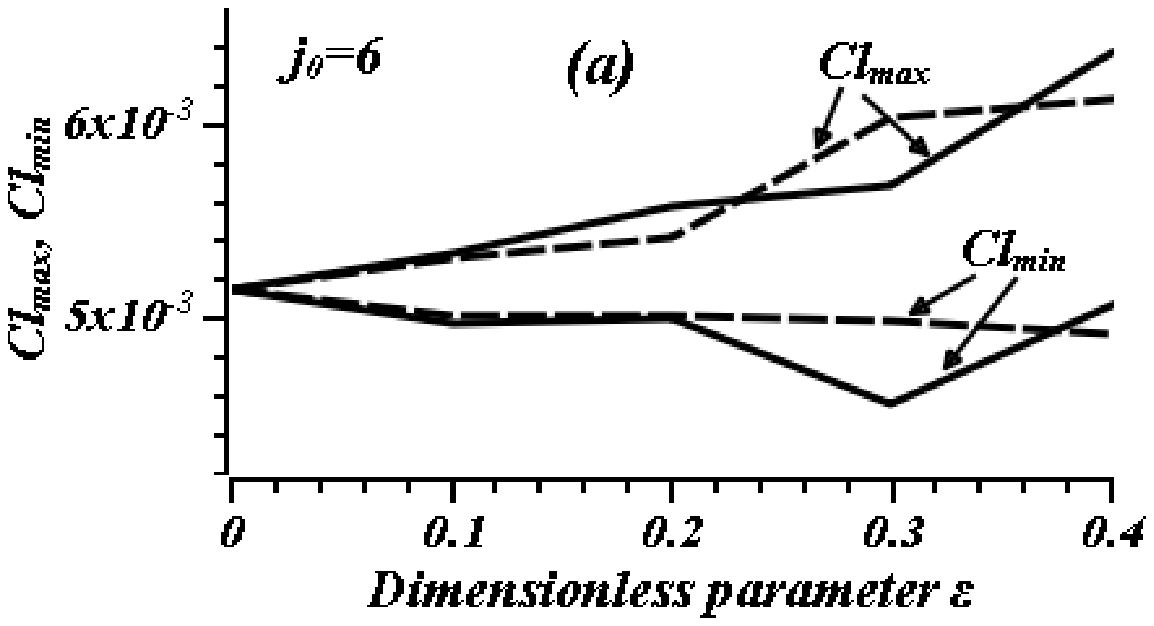,
  scale=0.7
   ,angle=0}
 \epsfig{file=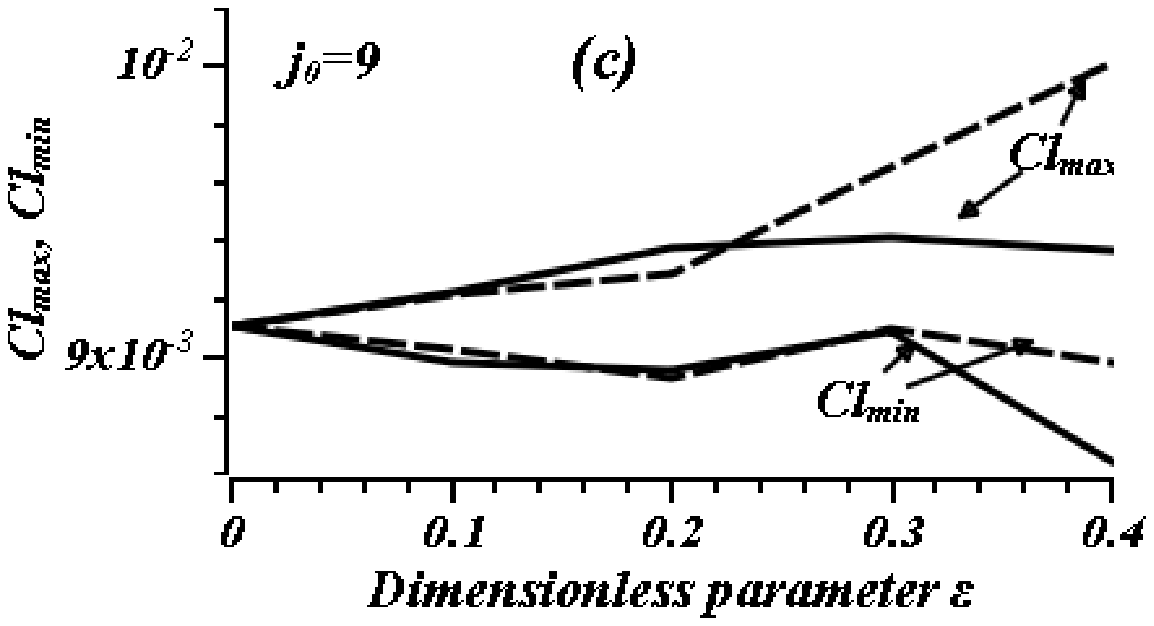, scale=0.7
   ,angle=0}
   \epsfig{file=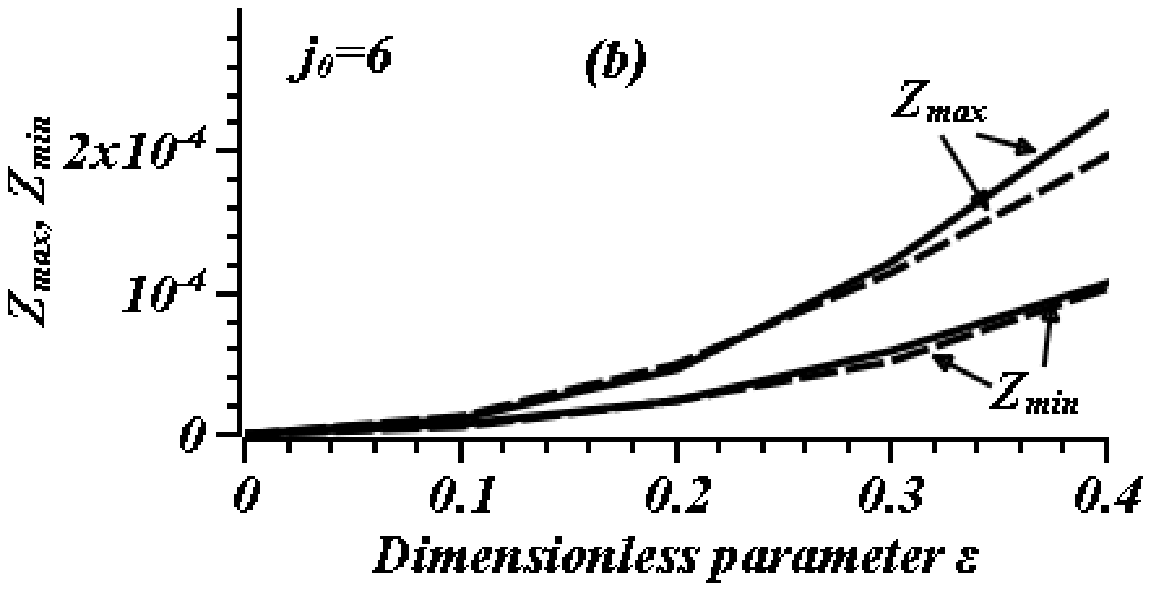,
  scale=0.7
   ,angle=0}
 \epsfig{file=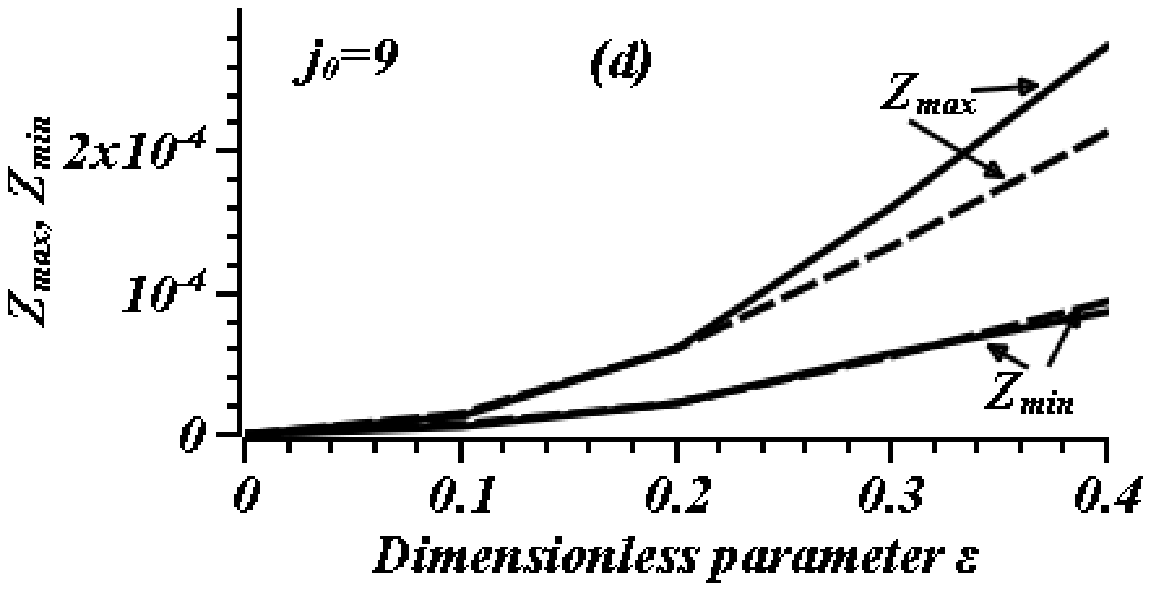, scale=0.7
   ,angle=0}
  \caption{Deformations of the fermion cluster in the system of $N=17$ 
  fermions with the noise-polarization of the initial state. 
 The pairwise
  discord is averaged over $10^2$ realizations of the random choices of  
  the parameters $\tilde b_j$, $j=1,\dots,17$, for each fixed value of the small parameter 
  $\epsilon$,
  $\epsilon=0,\;0.1,\;0.2,\;0.3,\;0.4$.
  The  cluster deformation by  noise effects
  for the density matrix  $\rho_1$ (solid lines) and $\rho_2$ (dashed lines) is characterized by 
  the functions $Cl_{max}(\epsilon)$, $Cl_{min}(\epsilon)$,
  $Z_{max}(\epsilon)$, $Z_{min}(\epsilon)$.
  ($a$) The initially polarized node  $j_0=6$, the functions $Cl_{max}(\epsilon)$ and 
$Cl_{min}(\epsilon)$. ($b$) The initially polarized node  $j_0=6$, 
the functions $Z_{max}(\epsilon)$ and 
$Z_{min}(\epsilon)$; ($c$) The initially polarized node  $j_0=9$, 
the  functions $Cl_{max}(\epsilon)$ and 
$Cl_{min}(\epsilon)$; ($d$) The initially polarized node $j_0=9$, 
the    functions $Z_{max}(\epsilon)$ and
$Z_{min}(\epsilon)$.} 
  \label{Fig:noise} 
\end{figure*}

Again, to characterize the deformation of the discord distribution caused by the noise-polarization,
we use the functions $Cl_{max}(\epsilon)$, 
$Cl_{min}(\epsilon)$ and $Z_{max}(\epsilon)$, 
$Z_{min}(\epsilon)$  defined by formulas (\ref{ClCl}) and (\ref{ZZ}) in which  we  replace $b$ with
$\epsilon$ to characterize the spread of the discord in the cluster $Cl$ and the spread 
of the parasitic discord, which is zero in the absence of  noise.
Pairs of   functions $Cl_{max}(\epsilon)$, 
$Cl_{min}(\epsilon)$  and  $Z_{max}(\epsilon)$, 
$Z_{min}(\epsilon)$ are shown, respectively, in Figs.\ref{Fig:noise}a and
$\ref{Fig:noise}b$  for $j_0=6$ and
 in Figs.\ref{Fig:noise}c and \ref{Fig:noise}d for $j_0=9$. We see 
 that the difference between the discord distribution 
 corresponding to the density matrices $\rho_1$ and 
 $\rho_2$ is not significant inside  the interval  
 $0\le \epsilon \le 0.4$, as is shown in Figs.\ref{Fig:noise}. More exactly, the curves 
 $Cl_{max}$ corresponding to the maximal discord for the density matrices $\rho_1$ and $\rho_2$ are 
 close to each other (see the upper solid and dashed lines in Figs.\ref{Fig:noise}a and \ref{Fig:noise}c),
 as well as the appropriate curves $Cl_{min}$  (see the lower solid and dashed lines in Figs.\ref{Fig:noise}a 
 and \ref{Fig:noise}c). The same statement holds for the curves  $Z_{max}$ and $Z_{min}$ in
 Figs.\ref{Fig:noise}b 
 and \ref{Fig:noise}d. 
  This serves us as a justification of 
 using the perturbation theory. 

\begin{figure*}
   \epsfig{file=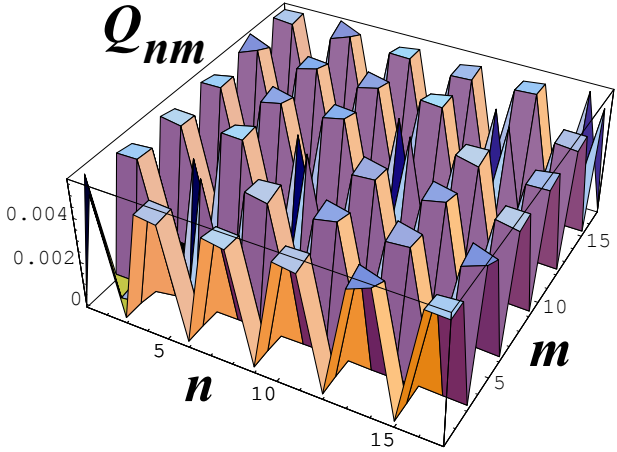,
  scale=0.8
   ,angle=0
}
  \caption{
  The averaged discord distribution $Q_{nm}$ in the  system of $N=17$ fermions with
  the initially polarized node 
  $j_0=6$, density matrix
  $\rho_2$ and  noise amplitude  $\epsilon=0.4$.
  The  pairwise  discord  is averaged over $10^2$ realizations of  random choices of  
  the parameters $\tilde b_j$, $j=1,\dots,17$. This distribution is slightly deformed 
  in comparison with that 
  in  Fig.\ref{Fig:3Dgen}a for $\epsilon=0$. The difference is visible in the peaks.
} 
  \label{Fig:3Dnoise} 
\end{figure*}

 To demonstrate the value  of the deformation of the  discord distribution  under the 
 small amplitude noise-polarization
 we represent the discord distribution
 in the cluster of correlated fermions for the initially polarized node 
$j_0=6$, the density matrix $\rho_2=\frac{\tilde \rho_2}{Z_2}$,  and the noise-amplitude
$\epsilon=0.4$  in Fig.\ref{Fig:3Dnoise}. The comparison of Fig.\ref{Fig:3Dnoise} with
Fig.\ref{Fig:3Dgen}a shows that 
the deformation 
of the discord distribution is approximately negligible in the cluster 
of the correlated fermions $Cl$. It is important that the noise does not destroy the 
stationarity of the discord distribution, similar to the parasitic polarization considered in 
Sec.\ref{Section:chain}.

\subsection{Stationarity of the pairwise discord in a fermion system with  noise}

In the previous section, we demonstrate that the pairwise discord in the Jordan-Wigner
fermion system with a single initially polarized node 
remains stationary under perturbations of two types: the parasitic polarization of
two neighboring nodes and the noise polarization considered by the perturbation method. In both cases, the 
 density matrix operator involves  at most three-fermion terms, see eqs.(\ref{rhot0e}) and
(\ref{rhotn}). 
However, it can be readily shown that the stationarity may not be destroyed by the noise 
polarization even if we take into account all 
terms of the perturbed density matrix. Or, even more generally, the distribution of the pairwise  discord
is stationary for the initial density matrix of the form
\begin{eqnarray}\label{inst1}
&&
\rho_0 =\frac{\tilde \rho}{Z}, \;\;Z=Tr\; \tilde \rho, \\\nonumber
&&
\tilde \rho= 1 +\sum_{i } \gamma_i I_{zi} + \sum_{i_1 \neq i_2 } \gamma_{i_1i_2} I_{zi_1}I_{zi_2}+
\sum_{i_1 \neq i_2 \neq i_3 } \gamma_{i_1i_2i_3} I_{zi_1}I_{zi_2}I_{zi_3}+ \dots+
\gamma_{1\dots N} I_{z1}\dots I_{zN},
\end{eqnarray}
where $\gamma$'s are scalar constants.
The evolution of the density matrix described by the Liouville equation 
$\frac{d\rho}{dt}=-i[H,\rho]$ reads:
\begin{eqnarray}\label{rho_t0}
\rho(t)= e^{-i t H} \rho_0 e^{i t H}.
 \end{eqnarray}
 After some transformations using equations (\ref{Hdiag}), (\ref{betc}), (\ref{Izc}) and 
(\ref{identity}), we obtain the density matrix in the form (we 
write the $t$-dependence explicitly)
\begin{eqnarray}\label{rhot0}
\rho(t) &=&\frac{1}{Z}\left(1 + \sum_{i=1}^N \alpha^{i}_{kk'} e^{-i t (\varepsilon_k -\varepsilon_{k'})}
\beta^+_k \beta_{k'} +\right.\\\nonumber
&&\left.
\sum_{i_1,i_2=1}^N \alpha^{i_1i_2}_{k_1k_2k_1'k_2'}
e^{-i t (\varepsilon_{k_1}+\varepsilon_{k_2}  -\varepsilon_{k_1'}-\varepsilon_{k_2'})}
\beta^+_{k_1} \beta_{k_1'}\beta^+_{k_2} \beta_{k_2'} + \dots \right),
\end{eqnarray}
where $\alpha$'s are expressed in terms of $\gamma$'s.
Eq.(\ref{rhot0}) is an infinite series. An important fact of its structure is 
that the product of the
operators 
$\beta_k^+\beta_{k'}$ appears together with the exponential  
$e^{-i t (\varepsilon_{k}-\varepsilon_{k'})}$. 
Considering the reduced  density matrix operator with respect to all fermions except for
the $n$th and $m$th ones  we obtain the density matrix in the form
\begin{eqnarray}\label{rhot000}
\rho_{nm}(t)= \tilde\alpha^{nm}_0 + 
\sum_{k,k'=n,m} \tilde \alpha^{nm}_{kk'} 
e^{-it(\varepsilon_k - \varepsilon_k')} \beta^+_k\beta_{k'} + 
  \tilde\alpha^{nm}_{nmnm}  \beta^+_{n}\beta^+_{m}\beta_{m}\beta_{n} ,
\end{eqnarray}
where all the coefficients $\tilde \alpha$'s  are  expressed in terms of the coefficients $\alpha$'s 
of eq.(\ref{rhot0}) and do not depend on $t$. We do not give the 
explicite expressions for the $\tilde \alpha$'s.
Terms of higher degrees in the $\beta$-operators do not appear in the two-particle density 
matrix operator (\ref{rhot000}) because of  the fermion operator property 
$\beta_k^2=(\beta_k^+)^2 =0$.
Using  the basis (\ref{basis}), 
we may represent the density operator (\ref{rhot000}) in the matrix form:
\begin{eqnarray}\label{rednm}
\rho_{nm}(t)=
\left(
\begin{array}{cccc}
\tilde \alpha^{nm}_0 &0&0&0\cr
0&\tilde \alpha^{nm}_0 +\tilde \alpha^{nm}_{nn}& e^{-it (\varepsilon_n -\varepsilon_m)}
\tilde \alpha^{nm}_{nm} &0\cr
0&e^{it (\varepsilon_n -\varepsilon_m)}
(\tilde \alpha^{nm}_{nm})^*& \tilde \alpha^{nm}_0 +\tilde \alpha^{nm}_{mm} &0\cr
0&0&0&\tilde \alpha^{nm}_0 +\tilde \alpha^{nm}_{nn}+\tilde \alpha^{nm}_{mm}+
\tilde \alpha^{nm}_{nmnm} 
\end{array}
\right).
\end{eqnarray}
Thus, the 
$t$-dependence appears only in the exponents in the non-diagonal elements.

Now we repeat the arguments used in the demonstration of the discord stationarity in 
Secs.\ref{Section:stationarity} and \ref{Section:noise}.  
Namely, it is shown in the Appendix, that the pairwise discord in  X-matrix
(\ref{rednm})
depends on the absolute value $|\tilde \alpha^{nm}_{nm}|$ of the non-diagonal element of this matrix.
Consequently  the discord does not depend on the time $t$.

%%%%%%%%%%%%%%%%%
\section{Conclusions}
\label{Section:conclusions}
In this paper we show that the  property of stationarity for 
the pairwise discord in the system of  
Jordan-Wigner fermions is stable with respect to  the polarization-like perturbations of the initial state. 
Two types of such parasitic polarizations are considered in details. The first one is associated with the 
experimental error in the 
creation of the single-node polarization initial state resulting in
low polarizations of the  neighboring nodes. 
The second one is related with the noise-polarization of all nodes. The only effect of both
such perturbations 
is deformation of the 
pairwise discord distribution in the cluster $Cl$ of  correlated fermions. In particular, 
such perturbations can destroy the cluster, which is 
explicitly demonstrated in Sec.\ref{Section:chain} for the case of two neighboring node
parasitic polarization. Thus, the discord stationarity in the Jordan-Wigner 
fermion system can be taken as a reliable and stable
advantage of the considered  fermion system in comparison with the original spin system. 
This encourages us to consider the possibility of a quantum gate realization on the basis 
of such systems of virtual particles.

This work is supported by the Program of the Presidium of RAS No.8 ''Development
of methods of obtaining chemical compounds and creation of new materials'' and
by the Russian Foundation for Basic Research,   grant 
No.13-03-00017.

%%%%%%%%%%%%%%%%%%%%%%%%%%%%
\section{Appendix: Quantum discord in the $X$-type state (\ref{rednm})}
\label{Section:appendix}
The two-particle density matrix  considered in 
this paper is a particular case of the so-called X-matrix \cite{ARA}:
\begin{eqnarray}\label{Xmatr}
\rho^{red}\equiv \rho^{(nm)}=\left(
\begin{array}{cccc}
\rho_{11} &0&0&0\cr
0&\rho_{22}&\rho_{23}&0\cr
0&\rho_{23}^*&\rho_{33}&0\cr
0&0&0&\rho_{44}
\end{array}
\right),\;\;\sum_{i=1}^4 \rho_{ii}=1.
\end{eqnarray}
The discord for  the X-matrix was studied in 
 \cite{ARA}. 
 Remind that  the discord between   particles $n$ and $m$  of a biparticle 
quantum system 
 can be calculated as 
 \begin{eqnarray}\label{Q}
Q_m={\cal{I}}(\rho) -{\cal{C}}^m (\rho),
\end{eqnarray} 
provided that the von Neumann type measurements are performed  over the particle
$m$. 
Here   ${\cal{I}}(\rho)$ is the total mutual information \cite{OZ}, which may be
written as follows: 
\begin{eqnarray}\label{I}
&&
{\cal{I}}(\rho) =S(\rho^{(n)}) + S(\rho^{(m)}) + \sum_{j=0}^3 \lambda_j \log_2
\lambda_j,\\\nonumber
\end{eqnarray}
where  $\lambda_j$ ($j=0,1,2,3$) are  the non-zero eigenvalues of the density
matrix $\rho^{(nm)}$,
\begin{eqnarray}
&&\lambda_0= \rho_{11}, \;\; \lambda_1= \rho_{44}, \\\nonumber
&&
\lambda_{2,3}=\frac{1}{2} \left(\rho_{22}+\rho_{33}\pm\sqrt{(\rho_{22}-\rho_{33})^2 + 
4 |\rho_{23}|^2 }\right),
\end{eqnarray}
and $\rho^{(n)}={\mbox{Tr}}_m \rho^{(nm)}$ and $\rho^{(m)}={\mbox{Tr}}_n
\rho^{(nm)}$ are the marginal density matrices.
 The appropriate entropies $S(\rho^{(n)})$ and $S(\rho^{(m)})$ are given by
the following formulas:
\begin{eqnarray}\label{SAB}
&&S(\rho^{(n)})=-(\rho_{11} +\rho_{22} ) \log_2(\rho_{11} +\rho_{22}) -
 (\rho_{33} +\rho_{44} ) \log_2(\rho_{33} +\rho_{44})
,\\\nonumber
&&S(\rho^{(m)})=-(\rho_{11} +\rho_{33} ) \log_2(\rho_{11} +\rho_{33}) -
 (\rho_{22} +\rho_{44} ) \log_2(\rho_{22} +\rho_{44}).
\end{eqnarray}
The so-called classical counterpart ${\cal{C}}^B (\rho^{(nm)})$ of the mutual
information 
can be found considering the minimization over  projective measurements performed 
on
the particle $m$ as follows \cite{ARA}:
\begin{eqnarray}\label{CB2}
&&
{\cal{C}}^{(m)} (\rho)=S(\rho^{(n)}) -\min\limits_{k\in[0,1]}(p_0 S_0 + p_1
S_1),
\end{eqnarray}
where
\begin{eqnarray}\label{S}
&&S(\theta_i)\equiv S_i = -\frac{1-\theta_i}{2}\log_2\frac{1-\theta_i}{2}-
%\\\nonumber
%&&\hspace{1cm}
                 \frac{1+\theta_i}{2}\log_2\frac{1+\theta_i}{2},
\\\label{p}
&&
p_0=(\rho_{11}+\rho_{33}) k + (\rho_{22}+\rho_{44})l, \;\; 
p_1=(\rho_{11}+\rho_{33}) l + (\rho_{22}+\rho_{44})k,\\\label{theta}
&&
\theta_0=\frac{1}{p_0}\sqrt{((\rho_{11}-\rho_{33})k + (\rho_{22}-\rho_{44})l)^2 + 4 kl |\rho_{23}|^2},\\\nonumber
&&
\theta_1=\frac{1}{p_1}\sqrt{((\rho_{11}-\rho_{33})l + (\rho_{22}-\rho_{44})k)^2 + 4 kl |\rho_{23}|^2}.
\end{eqnarray}
Here, the parameters $k$ and $l$ are related by the equation  \cite{ARA}
\begin{eqnarray}
k+l=1.
\end{eqnarray}
It is simple to show that the quantum discord $Q_n$ obtained performing 
 the von Neumann type measurements  on the particle $n$ is related with $Q_m$ 
as follows:
\begin{eqnarray}\label{QA}
Q_n=Q_m|_{\rho_{22}\leftrightarrow \rho_{33}}
\end{eqnarray}
for the system with the density matrix $\rho^{red}$ given by eq.(\ref{Xmatr}).
Then we define the  discord $Q_{nm}$ as the minimum of $Q_{n}$ and $Q_{m}$
\cite{FZ} 
\begin{eqnarray}\label{def_discord}
Q_{nm}=
\min(Q_n,Q_m), \;\; n\neq m
\end{eqnarray}
with the obvious property $Q_{nm}=Q_{mn}$.
We see that, if  $\rho_{nn}$, $n=1,2,3,4$,  and $|\rho_{23}|$ do not depend on the time $t$, then
the discord does not evolve with the time as well.

\end{document}